\documentclass[twocolumn,showpacs,preprintnumbers,amsmath,amssymb]{revtex4}
\usepackage{graphicx}
\usepackage{dcolumn}
\usepackage{bm}

\begin{document}

\title{Theoretical aspects of simple and nested
Fermi surfaces for superconductivity in
doped semiconductors and high-$T_C$ cuprates.}

\author{T. Jarlborg}

\affiliation{
DPMC, University of Geneva, 24 Quai Ernest-Ansermet, CH-1211 Geneva 4,
Switzerland. E-mail: thomas.jarlborg@unige.ch.
}


\begin{abstract}

The density-of-states at the Fermi energy, $N(E_F)$, is low in doped superconducting semiconductors
and high-$T_C$ cuprates.
This contrasts with the common view that
superconductivity requires a large electron-boson coupling
$\lambda$ and therefore also a large $N(E_F)$. However, the generic Fermi surfaces (FS)
of these systems are relatively simple.
 Here is presented arguments showing that
going from a 3-dimensional multi-band FS to a 2-dimensional and simple FS is energetically
favorable to superconductivity. Nesting and few excitations of bosons compensate for a low $N(E_F)$. 
The typical behavior of the 2-dimensional FS for cuprates, and small 3-dimensional FS pockets in
doped semiconductors and diamond, leads to $T_C$ variations as a function of doping in line with what
has been observed. Diamond is predicted to attain higher $T_C$ from
electron doping than from hole doping, while conditions for superconductivity in Si and Ge
are less favorable. A high-$T_C$ material should ideally
have few flat and parallel FS sheets with a reasonably large $N(E_F)$.

\end{abstract}


\maketitle


A. Cuprates;

D. Band structures;

D. Superconductivity.

\section{Introduction.}

Electron-phonon coupling, $\lambda$, is the likely cause of
superconductivity in
most, what now is called "low-$T_C$" superconductors like some elementary metals,
transition metal nitrides and carbides, A15-compounds etc. The 
superconducting $T_C$ has been described quite successfully by the BCS equation \cite{bcs},
or by the modified strong coupling McMillan form \cite{mcmill}. 
The estimates of $T_C$ along these lines often
show a reasonable correlation with measured $T_C's$ among many of the
low-$T_C$ materials, \cite{gomer,papa1,papa2,arb}. Usually, a large $\lambda$ needs
a large electronic density-of-states (DOS) at the Fermi energy, $E_F$.
However, many materials, in particular the high-$T_C$
cuprates and pnictides have low DOS, and cannot 
be understood from these approaches. It is remarkable that many new superconductors
have been discovered in the last two decades.
 The observation of superconductivity  
in near insulators like doped diamond \cite{diam}, and in 
weakly doped semiconductors, such as sodium doped WO$_3$  \cite{raub} or
Nb-doped SrTiO$_3$ \cite{shoo}, is surprising because 
the DOS  is low in these systems.
Several of the recently discovered superconductors are exotic in the sense that superconductivity
coexists with magnetism, as in heavy Fermion f-electron systems \cite{steg} and 
in Fe under pressure \cite{shim,jac}.
Spin-fluctuations are often present
in nearly ferromagnetic (FM) or anti-ferromagnetic (AFM) materials 
and might be a bosonic "glue" for superconductivity \cite{berk,fay,vanad,zrzn2}, perhaps
even in the high-$T_C$ cuprate systems \cite{pines,ueda}.
The existence of a pseudogap 
below a temperature $T^* > T_C$ in hole doped cuprates is now well established, as well as stripes and the evolution
of the Fermi surface (FS) as function of doping  \cite{tall,timu,dama}.
AFM stripe-like modulations on the Cu sub-lattice, with coupling to phonon distortions,
provoke a pseudogap in band calculations \cite{tj7}. 
The DOS at $E_F$, $N(E_F)$, and $\lambda$, in the cuprates is smaller than what is
typical for many low-$T_C$ metals and compounds. 
However, the
FS is strikingly simple in the cuprates with a cylindrical 2-dimensional (2D) shape \cite{dama}. This
is in contrast to the complicated 3-dimensional (3D) multi-structured FS's in transition metals and
compounds like the A15 superconductors \cite{jmp}. 

Here we examine why a simple
low-dimensional FS can be favorable to superconductivity and compensates for a low $N(E_F)$. Theoretical
considerations are described in sec. II. The results in sec. III are all based on band structures calculated with
self-consistent Linear Muffin-Tin Orbital (LMTO) method \cite{lmto} in the local-density approximation
(LDA) \cite{lda}.

\section{Theory.}

The weak coupling BCS formula is based on electron-phonon coupling and FE
bands.
\begin{equation}
k_BT_C = 1.13 \hbar\omega e^{-1/\lambda}
\label{eqbcs}
\end{equation}
where $\lambda$, the electron-phonon
coupling constant, often is calculated in density functional (DF) band calculations as \cite{mcmill,gomer,papa1,papa2,arb}
\begin{equation}
\lambda = N(E_F) I^2 / M \omega^2
\label{eqlamb}
\end{equation}
where $M$ is
an atomic mass, $\omega$ is the averaged phonon frequency 
and $I$ is the matrix element $< \partial V / \partial u >$, the change in electron
potential $V(r)$ due to the displacement $\partial u$. 
The denominator in eq. \ref{eqlamb} can also be written as
a force constant $K = \partial^2E/\partial u^2$, the second derivative of the total energy, $E$, with 
respect to the atomic displacement.
In the following we assume that $\lambda$ is coming from coupling to phonons only, but coupling
to spin-fluctuations can be considered from complementary corrections in the development \cite{tjfe,tj1}.

\begin{figure}
\includegraphics[height=6.0cm,width=8.0cm]{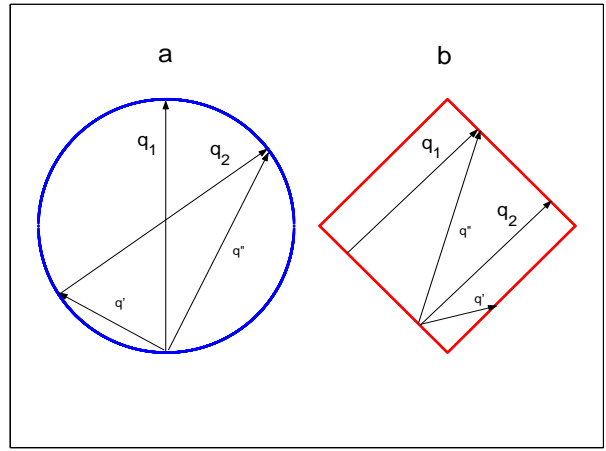}
\caption{Schematic 2D-projected FS in underdoped (a) and optimally doped (b) cuprates. Two different phonons
with vectors $q_1$ and $q_2$ are needed to open a gap on two small sections of the circular FS in case $a$.
For the square-like FS (case $b$) $q_1$ and $q_2$ are identical and one phonon is sufficient for opening of a gap
on the entire side of the FS. Phonon vectors $q'$ and $q"$ are less important
both in case $a$ and $b$, since they
do not span parallel sheets of the respective FS's. Case (a) can also represent a projection of a 3D spherical FS. 
}
\label{figcirc}
\end{figure}

The FS for the 3D FE band is a sphere with radius $k_F$, see Fig. \ref{figcirc}a. 
A phonon distortion ($u$) perturbs the lattice potential by $V_u(\vec{r}) = V_u exp(i \vec{q} \cdot \vec{r})$,
and a "gap" is created at new zone boundaries ($\pm q$) of the FE band \cite{zim};
the electronic states $\epsilon_k \rightarrow \epsilon_k \pm V_u$ at $|k|=q$.
This also holds for more general bands like the band crossing $E_F$ in the cuprates, where
supercell calculations with phonon distortions of the correct period induce gaps at $E_F$ \cite{tj7}. 
Any non-FE wave function can be written

\begin{equation}
\Psi^k(r) = 
e^{-i \vec{k} \cdot \vec{r}} \sum_G A_G e^{-i \vec{G} \cdot \vec{r}}
\label{eqpsi}
\end{equation}

where the last sum over large G-vectors describes the short-range wiggling of the wave function within the atoms.
This part will not be affected by $V_u(\vec{r})$, since the latter is a long-range
(longer than the atomic size) modulation in real space. Thus, $|q| < |G|$ and only the Bloch factor 
in eq. \ref{eqpsi} will be the modulation of the potential, as for FE bands.

The system gains energy only from states near $E_F$ (and the occupations at $E_F \pm V_u$),
and the FS is determining for what
phonons are involved. Thus a simple or nested FS helps superconductivity. 
This result might seem trivial,
but it points out that the often used
average of the entire phonon spectrum in eq. \ref{eqlamb} is not always appropriate in calculations of $T_C$.

After summing over all states $k,k'$ on the FS one finds an effective matrix element $I$ for optimal energy gains.
Nesting between parallel sheets of the FS makes up the
dominant contribution to the sum, i.e. for $\vec{q}=\vec{k}-\vec{k'}$ on the opposite side of the FS for 
FE bands, $\vec{k'} = -\vec{k}$. If less important non-nested vectors, like $\vec{q'}$ and $\vec{q"}$ in Fig. \ref{figcirc},
are neglected one has approximately that only phonons with $|q| = 2 k_F$ will contribute to $I$.


The coupling for harmonic vibrations is independent of amplitude ($u$), and it is
convenient to write $\lambda = N V_q^2 / Ku^2$, where $V_q$ is the matrix element
$< k~\Delta V~k'>$ for an excited phonon q. The numerator is the gain in electronic energy,
and the denominator is identified as the total cost of
elastic energy of the involved phonon spectrum.  An average over all phonon frequencies
up to the highest Debye frequency $\omega_{max}$ is; 

\begin{equation}
<\Omega> = \int^{\omega_{max}}_0 \omega F(\omega) \partial\omega 
\label{eqOmega}
\end{equation}
where $F(\omega)$ is the phonon DOS. This is reduced by a factor $f$ 
if only phonons between $\omega_1$ and $\omega_2$
are excited:
\begin{equation}
f = \int^{\omega_2}_{\omega_1} \omega F(\omega) \partial\omega / <\Omega>
\label{eqf}
\end{equation}

Instead of lumping the vibrational and electronic energy ratio into
one parameter $\lambda$ it is now convenient to take the logarithm of the $T_C$ equation and write 
from eqns. \ref{eqbcs}-\ref{eqlamb} \cite{ssc};

\begin{equation}
M <\Omega>^2 = N I^2 ln (1.13\hbar \omega/ T_C) 
\label{eqlog}
\end{equation}

If $I$ is constant at all parts of the FS, and with $I=V/u$ we have alternatively;

\begin{equation}
 Ku^2 = N V^2 ln (1.13\hbar \omega/ T_C)
\label{eqlog2}
\end{equation}
which permits to separate the cost in total vibrational energy ($Ku^2$) of participating phonons from
the gain in electron energy.

An atomic 3D lattice introduces anisotropy in the phonon spectrum where phonons along
[1,0,0] can be slightly different from the [1,1,0] or [1,1,1] directions.
A phonon (q) and its FS piece (k,-k) opens the gap over a part of the DOS, $N_q$,
of the total $N$. Assuming that the 26 highest symmetry directions are representative
for a 3D lattice we get 26 individual gap equations 
for the 6 [1,0,0], the 12 [1,1,0] and the 8 [1,1,1] directions if the FS is a sphere,
or more if there are multiple FS of complicated shapes. If one phonon q-vector is spanning
several different pairs of FS it should be counted only once, and a case with a 2D FS as
in Fig. \ref{figcirc}b can be reduced to only 2 phonons spanning the entire FS.
In the end there is a balance between the total elastic energy and the total gain of
electronic energy. In the following we show several examples where the FS's are simple, so that 
only few phonons contribute to the total elastic energy, which therefore will be advantageous
to superconductivity. 

\section{Results.}

\subsection{Doped SrTiO$_3$ and WO$_3$.}

The first example is given by the low-$T_C$ superconductivity in weakly
electron-doped semiconductors Na$_x$WO$_3$ and Nb$_x$SrTi$_{(1-x)}$O$_3$  \cite{raub,shoo}. 
The 3D bandstructures of the undoped materials
show wide band gaps typical of semiconductors \cite{matt}, but they become metallic through Na or Nb impurity doping. 
Superconductivity is surprising here, since the DOS
and $\lambda$ are small. The electron density is less than 0.01 $el$ per formula unit, $f.u.$, 
at optimal $T_C$ \cite{shoo}, when from band theory $N(E_F) \sim 0.15 (eV \cdot cell)^{-1}$
and the standard $\lambda$ will hardly be larger than 0.02, which is insufficient for
a detectable $T_C$ at normal metallic screening conditions \cite{sto}. The rigid-band model describes correctly 
the band structure and the FS (cf. Fig \ref{figcirc}a) of Nb$_x$SrTi$_{(1-x))}$O$_3$, with small
electron pockets centered at $\Gamma$  \cite{sto}. 
A Debye spectrum, $\omega = c q$, where $c$ is the sound velocity, is appropriate for small-q phonons,
and $F(\omega) = 9 \omega^2 / \omega_D^3$ in 3D, where $\omega_D$ is the Debye cutoff at large q \cite{zim}.
Since the FS diameter ($2k_F$) is small, only low-energy phonons with $q < 2k_F$ will contribute, and
 
 \begin{equation}
<\Omega> = \int^{\omega_{F}}_0 \omega F(\omega) \partial\omega = 9 \omega_F^4/4 \omega_D^3
\label{eqOmegaD}
\end{equation}

This makes $\Omega$ very small, much smaller than if all phonons up to $\omega_D$ had to be excited
for having a gap over the FS pocket. 
Hence, $NI^2/M<\Omega>^2$
in eq. \ref{eqlog} can be sufficiently large for a reasonable $T_C$ despite the smallness of $N$.

Another seemingly unexplained fact is that if $x$ increases also $N(E_F)$ and the standard 
$\lambda$ increase, so $T_c$ should
go up with $x$. Instead, $T_c \rightarrow 0$ for larger doping \cite{shoo}. This can be understood from 
the behavior of $N$ and $\Omega$ as function of $x$;
The free-electron like DOS of SrTiO$_3$ $N(\epsilon)$  
increases as $\sqrt(\epsilon)$ for $\epsilon$, the energy relative to the bottom of the
conduction band, thus $N \sim E_F^{1/2}$ \cite{sto}. But phonons in eq. \ref{eqOmegaD}  
contribute up to $\omega_F = ck_F$, and
since $E_F \sim k_F^2$ this makes $\Omega \sim E_F^2$. Thus, the advantage of a low $\Omega$ at low
doping disappears if $E_F$ increases through electron doping when the ratio $N(E_F)/\Omega \sim E_F^{-3/2}$  
becomes smaller.

\subsection{Doped diamond.}

\begin{figure}
\includegraphics[height=6.0cm,width=8.0cm]{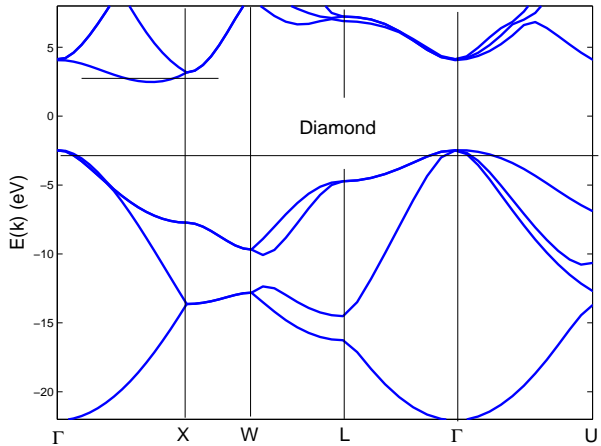}
\caption{The band structure of diamond along symmetry lines. The wide horizontal
line is the position of $E_F$ for a hole doping of 0.03 holes/atom. The short
horizontal line indicates $E_F$ for weak electron doping.
}
\label{figbnd}
\end{figure}

\begin{figure}
\includegraphics[height=6.0cm,width=8.0cm]{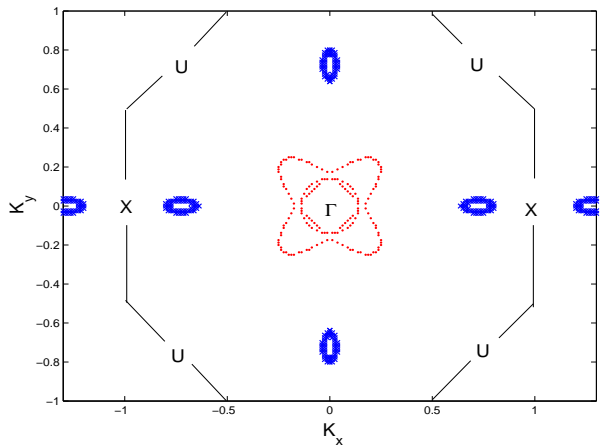}
\caption{Fermi surface plot in the $xy$-plane of hole- and electron-doped diamond. The FS  of hole doped diamond is made up by the $\Gamma$ centered
orbits shown by (red) dots. In electron doping the FS has small pockets indicated by (blue) x-points.
The scale for $K_x$ and $K_y$ are in units of 2$\pi/a_0$. 
}
\label{cdfs}
\end{figure}

A similar mechanism is applicable to boron-doped diamond, although it is now a question of
hole doping below the band gap. A $T_C \approx 4K$ was reported when the hole density is near 4.6 $10^{21} cm^{-3}$
\cite{diam}, or about 0.026 holes/atom. The explanation of this $T_C$ is often based on phonon
softening that accompanies the B doping together with a moderate $N(E_F)$ \cite{boeri,lee,bla,xia}. 
An inspection of the bands for hole doping as in Fig. \ref{figbnd} shows that the FS consists of 3 $\Gamma$
centered pieces, see Fig. \ref{cdfs}. Many phonon excitations are possible for $q < \sim 0.6$ of the $\Gamma-U$ distance. 
Umklapp transitions are possible for still larger $q$, and they become more frequent if the doping increases. 
The factor $f$ for reducing the phonon average is not as small as in section III-A, but
the situation is qualitatively the same, and the
benefits from few phonon excitations are lost quite soon if the doping
increases.

An interesting situation appears if electron-doping can be made in diamond. Fig. \ref{figbnd} shows the possibility
for an electron pocket at 3/4 of the $\Gamma-X$ distance if electron doping doping is made. 
Doped diamond is not an ionic material, and the matrix element $I$ is dominated by dipole scattering,
$\Delta \ell = \pm 1$, which can be calculated directly from the band structure
by use of the Rigid Muffin-Tin Approximation (RMTA) \cite{gomer,daco}
Such
calculations with rigid-band shifts of $E_F$ show that
the matrix element $NI^2$ is 40 percent larger for a doping of 0.03 electrons/atom than for 0.03 holes/atom,
see Table I.
This is partly due to a moderate amount of d-states in the upper band, which permits an enhanced p-d scattering.
The $\ell,m$-character of the electron-pocket state is otherwise quite similar to
the s-p states near the hole-pocket states at $\Gamma$. The electron-doped FS is made up by small pockets
on the 6 $\Gamma-X$ lines, as shown in Fig \ref{cdfs}. Small-$q$ scattering 
within these pockets is possible, but
with limited importance because of their small fraction of the
total $F(\omega)$. In addition, screening will diminish the matrix elements
for the very long wave lengths of these waves. 
Few large $q,\omega$ scatterings between the pockets ($q$-vectors of type $[\frac{1}{2},0,0]$ 
and $\sqrt(2)[\frac{3}{4},\frac{3}{4},0]$) are possible together with their Umklapp
scatterings. As a whole these are fewer $q$-excitations 
than for the hole-doped FS. Therefore, larger matrix elements together with fewer
phonon excitations suggest that electron doped diamond is more promising for a high $T_C$ than with hole doping. 
This assumes that substitutions of C with N can be made, and that rigid-band conditions prevail.

\begin{table}[ht]
\caption{\label{table1} Calculated energy gap $E_g$ (eV), DOS at $E_F$ (states/cell/eV) and
the matrix elements $NI^2$ (eV/\AA$^2$) for doping levels corresponding to 0.03 holes/cell
($hole$) and 0.03 electrons/cell ($electr.$) for C, Si and Ge.}
  \vskip 2mm
  \begin{center}
  \begin{tabular}{l c c c c c}
  \hline
     Element  & $E_g$  & $N(hole)$  & $NI^2 (hole)$ & $N(electr.)$ & $NI^2 (electr.)$ \\

  \hline \hline
  C  & 4.95 & 0.17 & 10.5 & 0.15 & 14.0 \\
  Si & 1.09 & 0.38 &  2.2 & 0.32 &  1.8 \\
  Ge & 0.76 & 0.31 &  1.8 & 0.39 &  2.0 \\
  \hline
  \end{tabular}
  \end{center}
  \end{table}

 One could suspect that also other electron-doped semiconductors 
of diamond structure are interesting
for superconductivity, at least if their matrix elements are comparable.
A similar conduction band minimum as in diamond exists in Si, with possible electron pockets
at about the same k-point between $\Gamma$ and $X$. In Ge the minimum is at the $L$-point and GaAs has a direct gap.
Electron doping in the latter system forms a continuous FS pocket around the $\Gamma$ point,
which invokes multi-directional $\vec{q}$-phonons with small amplitudes, as for hole-doped diamond. In the former
case the scattering is not only within $L$-pockets, but also for intermediate $q$-amplitudes between the pockets.
However, the calculated matrix elements are much lower in the small-gap semiconductors than for diamond, see Table I
for equivalent levels of hole and electron doping.
This is despite a more rapid increase of $N(E_F)$ as a function of both hole- and electron doping, $x$.
The latter is for Ge more promising than hole doping, as is concluded from the matrix elements, but
they are still too small in comparison to the $NI^2$-values for diamond to be interesting for enhanced
superconductivity. Small amounts of boron doping in Si are reported to have a $T_C \approx$ 0.35 K \cite{bust}. 

\subsection{High-$T_C$ cuprates.}

The FS's in high-$T_C$ cuprates are interesting for nesting. The cuprates are essentially 2D materials with
anisotropic phonons, and the
 FS is a cylinder oriented along $\vec{k_z}$
perpendicular to the CuO planes, as for case $a$ in Fig. \ref{figcirc}. They are more complicated than
the semiconductors because multiple phonon energies exist for the large unit cells. 
Phonons with $q$-vectors
along $\vec{z}$ are not much involved in the absence of nesting in this direction. Ideally only
phonons with $q$-vectors along the 
4 [1,0] and 4 [1,1] directions would contribute, which is already a large reduction compared to
the 3D lattice. Another profitable case is if a single phonon
would be able to open the gap at several FS-pieces. This will be the case when the 2D FS cylinder
distorts into a "diamond"-shaped FS, as it does in the cuprates when the hole doping is increased to
about 0.15 holes/Cu, see case "b" in Fig. \ref{figcirc}. 
The Fermi energy then reaches the DOS peak caused by the van-Hove singularity (vHS) and $N$ goes up, which
also increases $\lambda$ and $T_C$, but in addition; the cost in elastic energy is reduced further when
only 2 q-vectors (with multiple $\omega_q$) are scanning two sets of parallel FS sheets. This case at optimal doping 
is characterized by a reduced number of excited phonon/spinwave
excitations compared to the over- or under-doped cases. 
From the reduced number of phonons, $f < 1$, which translates into an increase in $\lambda$ by a factor of $1/f$ and
$T_C$ is boosted. 
But on the other hand, the effect of
phonon softening will be focused on those few modes. The term $Ku^2$ becomes negative, $\omega \rightarrow 0$,
and equations \ref{eqbcs},\ref{eqlog} and \ref{eqlog2} are no longer applicable. 
Softening and competing charge density waves are expected to limit superconductivity
in 1D organic systems with flat FS's \cite{edw}. For cuprates it is likely that the softening
can be tempered by spin-phonon coupling. The reason is that spin-wave excitation energies are usually higher
than phonon energies, so "in-phase" mixing of spin waves into a phonon will push the energy of the latter upwards.
However, long-range modulations can be a result of competing spin/phonon waves, which generate FS reconstructions
even if superconductivity survives \cite{harr,seb,tj11}.

 \begin{figure}
\includegraphics[height=6.0cm,width=8.0cm]{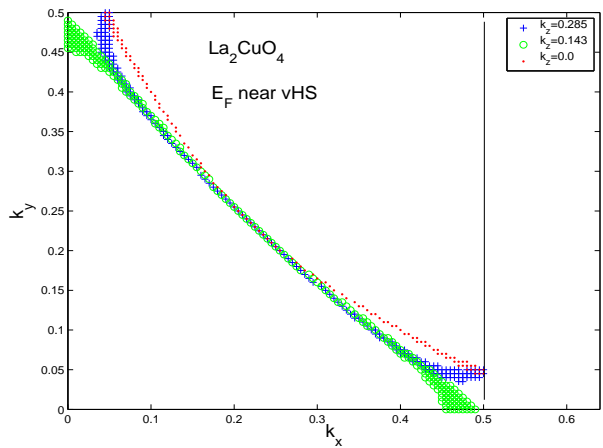}
\caption{The FS for 3 $k_z$-planes in La$_2$CuO$_4$ where $E_F$ is adjusted
5 mRy downwards in order to be at the van-Hove singularity corresponding to a
rigid-band doping of 0.12 holes per cell.
}
\label{figFSl}
\end{figure}

 \begin{figure}
\includegraphics[height=6.0cm,width=8.0cm]{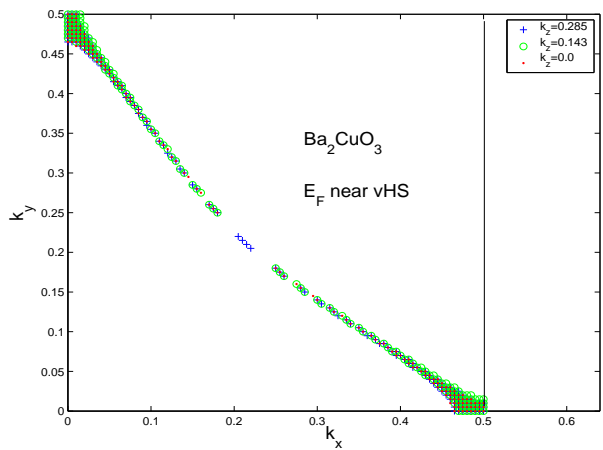}
\caption{The FS for 3 $k_z$-planes in Ba$_2$CuO$_3$ where $E_F$ is adjusted
8 mRy downwards in order to be at the van-Hove singularity corresponding to a
rigid-band doping of 0.17 holes per cell. Note that contrary to Fig. \ref{figFSl} the FS
is identical at the 3 levels of $k_z$.
}
\label{figFSb}
\end{figure}

As mentioned above,
the calculated FS's of undoped high-$T_C$ cuprates are essentially 2D and consist of M-centered barrels 
with a circular $k_z$-projection as in Fig. \ref{figcirc}a. It becomes straightened out and
reaches the X- and Y-points for increased
hole doping as the calculated one for La$_2$CuO$_4$ (LCO) shown in Fig. \ref{figFSl} for
a rigid-band doping of 0.12 holes per Cu. The Fermi level crosses the top of
the DOS peak at the vHS of the band. The $k_z$ dispersion is
small but visible by the FS's at three levels of $k_z$ in Fig. \ref{figFSl}. This is quite close to
the idealized nested FS of Fig. \ref{figcirc}b. For even higher hole doping, at "over doping", the FS
takes a barrel-like shape again, but now centered at the $\Gamma$-point. 
One idea for having a more
nested 2D FS (at an optimal hole doping when the 2D projected FS is rather straight and the band is near
the vHS) with less $k_z$ dispersion is to cut the orbital overlap along $\vec{z}$ between
Cu-d and apical O-p states. Removing apical oxygens would be one possibility, but it also increases
the electron doping according to calculations \cite{bce}. It is interesting, but not
fully understood, that O-deficient Ba$_2$CuO$_{(4-\delta)}$ (BCO) and Sr$_2$CuO$_{(4-\delta)}$ (SCO), 
which are of the same structure
as LCO, can have a much higher $T_C$ than LSCO \cite{liu,gao,geba}. Experimental and theoretical
works indicate that apical (and not planar) oxygens are missing in BCO \cite{liu,gao,bce}.
In fact, calculations show that Ba$_2$CuO$_3$ has a very similar electronic
structure to undoped LCO \cite{bce}. Moreover, the absence of one O-layer is expected to diminish
the already small orbital overlap along $\vec{z}$, and hence to make the FS even more uniform along $k_z$
than in doped LCO. This is indeed what happens; Fig. \ref{figFSb} shows the FS 
for Ba$_2$CuO$_3$ when $E_F$ is moved to coincide with the vHS at a hole doping of 0.17 holes/Cu.
(Note that the doping in LCO is made via La/Sr or La/Ba exchange, while in BCO and SRO the doping level
is determined by the missing O on apical positions \cite{bce}.)
The three FS cuts for different $k_z$ fall on top of each other, which is not the case for LCO.
The $N(E_F)$ at the vHS's of the two cases are comparable (23.5 and 25.2 states/cell/Ry for BCO
and LCO, respectively), and cannot explain the large difference in $T_C$ between the two
systems. Therefore, it is suggested that the almost perfect nesting of the FS for BCO
makes $T_C$ larger. Superconductivity requires fewer 
phonon and/or spin fluctuation excitations in that case. 

Many of the energy-costly non-nested
modes will be excluded if the wide diagonal FS in Fig. \ref{figFSb} can be shortened into a short bar
halfway between the end points. This is supposed to happen in 
cuprates with pseudogaps and stripes, when a potential
modulation removes states at the end points of
the flat FS, leaving only the mid-section of the FS behind \cite{tj7}.    
There are recent efforts to understand
the role of ordering of defects in the cuprates, since it also tends to open up pseudogaps near the
X- and Y-points of the FS \cite{bianc,apl}. 
 Quantitative results for $T_C$ require
precise evaluation of the energetics of nested as well as non-nested excitations. 

\section{Conclusion.}

This work shows that a high $T_C$ is
possible despite a small $N(E_F)$ if few phonon excitations
are required for the pairing in superconductors with simple FS. 
The ordinary $\lambda$ is an appropriate parameter only if the
FS is complicated and fills all parts of the Brillouin zone, when also $N(E_F)$ is rather large.
The occurrence of superconductivity in doped 3D semiconductors with very low $N(E_F)$ 
can be understood qualitatively from the 
smallness of the FS at the
zone center. 
The situation in hole-doped diamond is quite similar, while with electron doping a very
different FS appears. Fewer phonon excitations and larger matrix elements suggest higher $T_C$ 
with electron doping than with hole doping. 
It is tempting to suggest that weakly doped Si, Ge or GaAs could have a comparable $T_C$,
since their FS shapes are similar to that of doped diamond. However, the matrix elements are
significantly smaller.
For the layered cuprates it is possible that the 2D-shaped FS implies a large reduction
of the energy for phonon/spin excitations in the superconducting process, so that $T_C$ can be high
despite the modest $N(E_F)$. Very similar electronic structures in optimally doped La$_2$CuO$_4$
and Ba$_2$CuO$_{(4-\delta)}$, but less warping of the FS and much higher observed $T_C$ in the latter
support this hypothesis. It is suggested that very flat sections of the FS, or even truncated
"arcs", require a minimum of phonon/spin excitations.   
Not only the amplitude of $N(E_F)$, but also the FS shape and peaks in the generalized susceptibility
\cite{sorm}, should be considered when searching for good superconductors.

\end{document}